\documentstyle[aps,multicol,epsf]{revtex}
\input epsf
\begin{document}
\draft

\title{Finite Size and Current Effects on IV Characteristics of Josephson
Junction Arrays}

\author{M.V. Simkin and J.M. Kosterlitz}

\address{Department of Physics, Brown University,
Providence, RI 02912, U.S.A}

\date{\today}
\maketitle

\widetext

\begin{abstract}
The effects of finite size and of finite current on the current-voltage
characteristics of Josephson junction arrays are studied both theoretically
and by numerical simulations. The cross-over from non-linear to linear
behavior at low temperature is shown to be a finite size effect and the
non-linear behavior at higher temperature, $T>T_{KT}$, is shown to be a
finite current effect. These are argued to result from competition between
the length scales characterizing the system. The importance of
boundary effects is discussed and it is shown that these may dominate the
behavior in small arrays.
\end{abstract}

\begin{multicols}{2}
\narrowtext

Two dimensional ($2D$) Josephson junction arrays ($JJA$) in the absence of
an external magnetic field have been extensively studied
\cite{01,02,03,04,05,06,ls} as model systems to observe the
Kosterlitz-Thouless
(KT) transition\cite{07,08}. In the thermodynamic limit when the array size
$N\rightarrow\infty$ and the applied current $I\rightarrow 0$ such that
$IN\rightarrow\infty$ the current-voltage ($IV$) relation
is $V\sim I^{a(T)}$ where the exponent $a(T)\geq 3$ when the temperature
$T\leq T_{KT}$ and $a(T)=1$ when $T>T_{KT}$\cite{09,10}. However, it is
observed in experiments on small arrays\cite{01} and in simulations\cite{11}
that the $IV$ relation for $T<T_{KT}$ becomes linear when the
applied current $I$ is lowered below some threshold value. An explanation
that this is due to a residual magnetic field inducing some free vortices
was proposed\cite{01} and an opposing point of view that this is an
intrinsic effect was offered\cite{11}. However, experiments on large
arrays\cite{05} show no cross-over to linear behavior at $T<T_{KT}$ down to
the lowest accessible currents which implies that a low $T$ linear $IV$
relation is a finite size effect. In this article, we demonstrate both
analytically and numerically that this is a finite size effect and that the
nonlinear $IV$ relation becomes linear for $I<I_{x}\approx I_{o}/N$ where
$I$ is the applied current per junction and $I_{o}$ is the critical current.
It is also observed that the $IV$ relation is also non-linear at
temperatures slightly above $T_{KT}$ and we show that this is a finite
current effect for $I>I_{f}\approx I_{o}/\xi_{+}(T)$ where $\xi_{+}(T)$ is
the dimensionless correlation length for $T>T_{KT}$.

The behavior of the system is controlled by the length scales in the
problem: the linear size $N$ of the array, the current length
$\xi_{I}=I_{o}/I$ which is the maximum size of a bound vortex pair in the
presence of of a current $I$ and the thermal correlation length $\xi(T)$
which diverges as $T\rightarrow T_{KT}^{+}$ and is infinite for $T\leq
T_{KT}$. Above $T_{KT}$, $\xi =\xi_{+}(T)$ may be interpreted as the size of
the largest bound pair which is stable against thermal fluctuations, while
bound pairs of all sizes exist for $T\leq T_{KT}$. The essential difference
is that, for $T>T_{KT}$, there is also a finite density of thermally excited
free vortices leading to a linear resistivity $R\sim\xi_{+}^{-2}$ while for
$T<T_{KT}$ the only free vortices are due to the current unbinding of vortex
pairs which leads to a non-linear $IV$ relation. In the $2D XY$ model which
describes the system, there is a fourth length scale $\xi_{-}(T)$ which also
diverges at $T_{KT}$ and $\xi_{+}\sim\xi_{-}^{2\pi}$\cite{ahns}. This length
plays a
rather different role to the others as it is a measure of the scale at which
the renormalized vortex interaction $K(l)$ is essentially at its asymptotic
value $K(\infty)=K_{R}(T)$. A
renormalization group analysis shows that $\xi_{-}$ is the scale at which
deviations from critical behavior become significant and not the scale at
which the RG equations become invalid. The other scales $N,\xi_{+},\xi_{I}$
are scales at which the RG equations do become invalid and, to proceed, some
physically motivated approximation such as Debye-H\"uckel\cite{bn} must be
made and the result extrapolated back.

A scaling assumption for the resistivity $R$ which is consistent with the
renormalization group is
\begin{equation}
\label{eq.one}
V/I=e^{-zl}{\cal R}(Ne^{-l},\xi e^{-l},\xi_{I}e^{-l},l/ln\xi_{-})
\end{equation}
where the dynamical exponent $z=2$ and ${\cal R}$ is an unknown scaling
function. This scaling form is familiar from more conventional scaling
functions with the exception of the combination $l/ln\xi_{-}$ which is a
consequence of the marginally relevant and irrelevant scaling fields in the
system. The ultimate effect of this is to produce temperature and current
dependent power laws. One can obtain some information from the scaling
ansatz by choosing a value of $l$ at which the scaling function ${\cal R}$
can be computed. For example, in the case $\xi_{I}<N,\xi$, one can
choose $e^{l}=\xi_{I}$ at which scale all vortex pairs are unbound by the
current and the vortices may be regarded as moving independently in a
viscous medium driven by the applied current so that
\begin{equation}
V/I=I^{2}{\cal R}(N/\xi_{I},\xi/\xi_{I},1,ln\xi_{I}/ln\xi_{-})
\end{equation}
For $T\leq T_{KT},N\rightarrow\infty$ and $ln\xi_{I}>ln\xi_{-}$ we expect
\begin{equation}
{\cal R}(\infty,\infty,1,ln\xi_{I}/ln\xi_{-})\sim I^{1/ln\xi_{-}(T)}\sim
I^{x(T)}
\end{equation}
where $\xi_{-}(T)=exp(1/x(T))$ with $x(T)=b|T-T_{KT}|^{1/2}$, which leads to
the standard nonlinear $IV$ relation $V\sim I^{\pi K_{R}(T)+1}$ as the
renormalized stiffness constant $\pi K_{R}(T)=2+x(T)$ and,
when $ln\xi_{I}<ln\xi_{-}$, we expect $V\sim I^{3}$. Finite size dominated
behavior
is also contained in Eq.(\ref{eq.one}) by taking $N<\xi_{I}$ and $e^{l}=N$
to obtain
\begin{equation}
V/I=N^{-2}{\cal R}(1,\infty,\xi_{I}/N,lnN/ln\xi_{I})
\end{equation}
Now, when $lnN>\ln\xi_{-}$, identical arguments lead to a linear $IV$
relation $V/I\sim N^{-\pi K_{R}(T)}$ and $V/I\sim N^{-2}$ when
$lnN<ln\xi{-}$. By making appropriate assumptions about the behavior of the
unknown scaling function ${\cal R}$ in the various limits, all the expected
behaviors can be reproduced and one expects qualitative changes when one of
the ratios of lengths is of order unity. Of course, real dynamical
calculations should be done to determine the functional form of ${\cal R}$
at the scale $l$ and then use the RG equations to extrapolate to physical
values of the parameters. We have been unable to do this explicitly but hope
to address this in the future. The strategy of this paper is to numerically
simulate the $IV$ relation and to interpret the data by these scaling
considerations.

We have performed simulations on unfrustrated square $N\times N$ arrays of
various sizes from $N=4$ to $N=64$ at different temperatures to measure the
voltage across the array as a function of the external current $I$ per bond.
We use a modification of the Langevin dynamical method
\cite{06,ls} as proposed by Falo, Bishop and Lomdahl\cite{12}. We assume
that
all junctions have the same critical current $I_{o}$ and are shunted by equal
resistances $R$ Each superconducting grain at the vertices of the lattice
has a capacitance $C$ to ground. The dynamical equations for the phase
$\theta_{n}$ and the voltage $V_{n}$ of the superconducting grain at
site ${n}$ follow from charge conservation and the Josephson equation.
\begin{eqnarray}
\label{eq:a}
d\theta_{n}/dt &=& 2eV_{n}/\hbar       \nonumber \\
CdV_{n}/dt &=& I_{o}\sum_{<m>}sin(\theta_{m}-\theta_{n}) + \nonumber \\
R^{-1}\sum_{<m>}(V_{m}-V_{n}) + \sum_{<m>}I_{
mn}^{th}
\end{eqnarray}
The sum over $<m>$ is over the nearest neighbors of site $n$ and
$I_{mn}^{th}$ is a thermal noise current in the bond $mn$
satisfying the fluctuation dissipation relation
\begin{equation}
\label{eq:b}
<I_{ij}^{th}(t)I_{kl}^{th}(t')>
= (2T/R)(\delta_{ik}\delta_{jl}-\delta_{il}\delta_{jk})\delta(t-t')
\end{equation}
The capacitances $C$ to ground are introduced purely for computational
convenience and the simulations were performed with an intermediate value of
the McCumber-Stewart parameter\cite{ms} $\beta=2e^{2}R^{2}I_{o}C/\hbar =1$.

To imitate a typical experimental configuration, we use current injection
and extraction from superconducting busbars at two opposite edges of the
array. The busbars are modelled by columns of $N$ superconducting grains
coupled by infinitely strong junctions, each of these grains again having a
capacitanc $C$ to ground so that, in the absence of a magnetic field, the
phase $\theta$ and the potential $V$ will be the same everywhere on a bar.
The busbars are each connected to the array by a set of $N$ junctions which
are assumed identical to those in the array. With this geometry, separate
equations of motion are necessary for the phases $\theta_{L},\theta_{R}$ and
potentials $V_{L},V_{R}$ of the busbars on the left and right edges of the
array. Modelling the bars in this way, they are very similar to the
equations of motion for the phases and potentials of the sites in the
interior of the array.
\begin{eqnarray}
\label{eq:c}
d\theta_{L}/dt &=& 2eV_{L}/\hbar   \nonumber \\
d\theta_{R}/dt &=& 2eV_{R}/\hbar   \nonumber \\
CdV_{L}/dt &=& I+I_{o}N^{-1}\sum_{i=1}^{N}sin(\theta_{i}-\theta_{L}) +
\nonumber \\
R^{-1}N^{-1}\sum_{i=1}^{N}(V_{i}-V_{L})   \cr
CdV_{R}/dt &=& -I+I_{o}N^{-1}\sum_{i=1}^{N}sin(\theta_{i}-\theta_{R}) +
\nonumber \\
R^{-1}N^{-1}\sum_{i=1}^{N}(V_{i}-V_{R})
\end{eqnarray}
where $I$ is the current per bond injected into the left busbar and
extracted from the right. The sums over $i$ are over the $N$ sites connected
to the busbars.

This busbar geometry has some advantages over the more usual method of
uniform current injection where an equal current is injected or extracted
from each site at the edges of the array and is not technically difficult to
implement. In fact, it is easier in the presence of an external magnetic
field as the gauge invariant phase is constant along the busbars and the
currents in the bonds attached to them are automatically properly accounted
for, in contrast to uniform injection. The potential drop across a piece of
the array in the presence of an applied current is due to free vortices,
which may be produced by thermal excitation, by the applied current or by an
external magnetic field, being driven across the array perpendicular to the
current. Uniform injection emphasizes dissipation at the edges where the
current is injected and extracted as vortices are easily created
there and will be attracted to the edges by
their images. Unbinding of a vortex/image pair is not needed for dissipation
but vortices at the edges are driven across the array by the applied current
and, in a finite system, most of the dissipation may occur at the
edges and not in the bulk. This implies that there is a significant voltage
drop across narrow regions at the edges and that these boundary effects may
dominate in finite arrays. In the busbar geometry used in the simulations
reported here, vortices cannot be created adjacent to a superconducting
busbar and are repelled from the
array edges so that dissipation there will be suppressed. One thus expects
that
there will be a negligible voltage drop at the edges and all will occur in
the bulk of the array. We have done simulations in both geometries to
confirm this picture and find that, with the system sizes used, most of the
potential drop is at the edges with uniform injection.

\begin{figure}
\epsfxsize= 7 cm \epsfbox{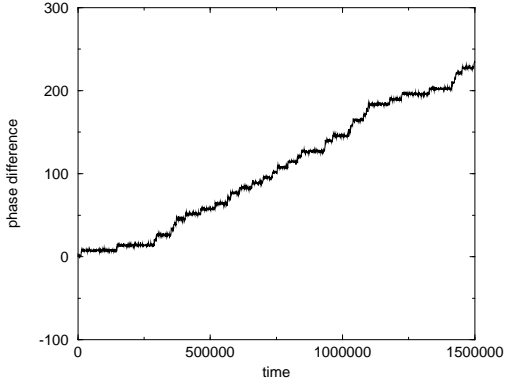}
\caption{Phase difference $\phi(t)$ across a $N=32$ array at $T=0.8$ and
$I/I_{o}=0.025$.}
\label{fig1}
\end{figure}

Simulations were performed on $N\times N$ square arrays in the busbar
geometry with $N=4,8,16,32,64$ at $T=0.8,1.0,1.1,1.3$ with the critical
temperature $T_{KT}\approx 0.89$\cite{olsson} where $T$ is in units of
$\hbar I_{o}/2e$. The equations of motion, Eqs.(\ref{eq:a}-\ref{eq:c}), were
integrated using the simple Euler algorithm with a time step $\Delta t=0.05$
in units of $1/\omega_{J}=(\hbar C/2eI_{o})^{1/2}$, the inverse Josephson
plasma frequency. Decreasing the time step by a factor of $10$ did not
change the results. The system is initially in a configuration at $t=0$ with
all $\theta_{n}=0=V_{n}$ and integrated over at least $10^{6}$ time steps
and the phase difference $\phi(t)\equiv\theta_{R}(t)-\theta_{L}(t)$ recorded
as a function of $t$ (see Fig.1). From this, it is clear that $\phi(t)$
grows by discontinuous jumps of magnitude $2\pi n$ due to the motion of
vortices. The mean voltage drop $V$ across the system is given by
\begin{equation}
\label{eq:v}
V/RI_{o}=(\phi(t_{r})-\phi(0))/t_{r}
\end{equation}
where $t_{r}$ is the run time in units of $1/\omega_{J}$. To estimate the
errors, we divide each run into four equal intervals and estimate $V$ for
each interval to obtain $\Delta V/V\approx 0.1$. Since the voltage $V$ is
caused by vortices crossing the array, the number $n_{v}$ of these is
\begin{equation}
\label{eq:nv}
2\pi n_{v}=\phi(t_{r})-\phi(0)
\end{equation}
The run time $t_{r}$ is chosen so that $n_{v}>100$ and one expects an error
in $V$ of $\Delta V/V\approx n_{v}^{-1/2}\leq 0.1$ which is consistent with
the estimate from block averaging.

\begin{figure}
\epsfxsize= 7 cm \epsfbox{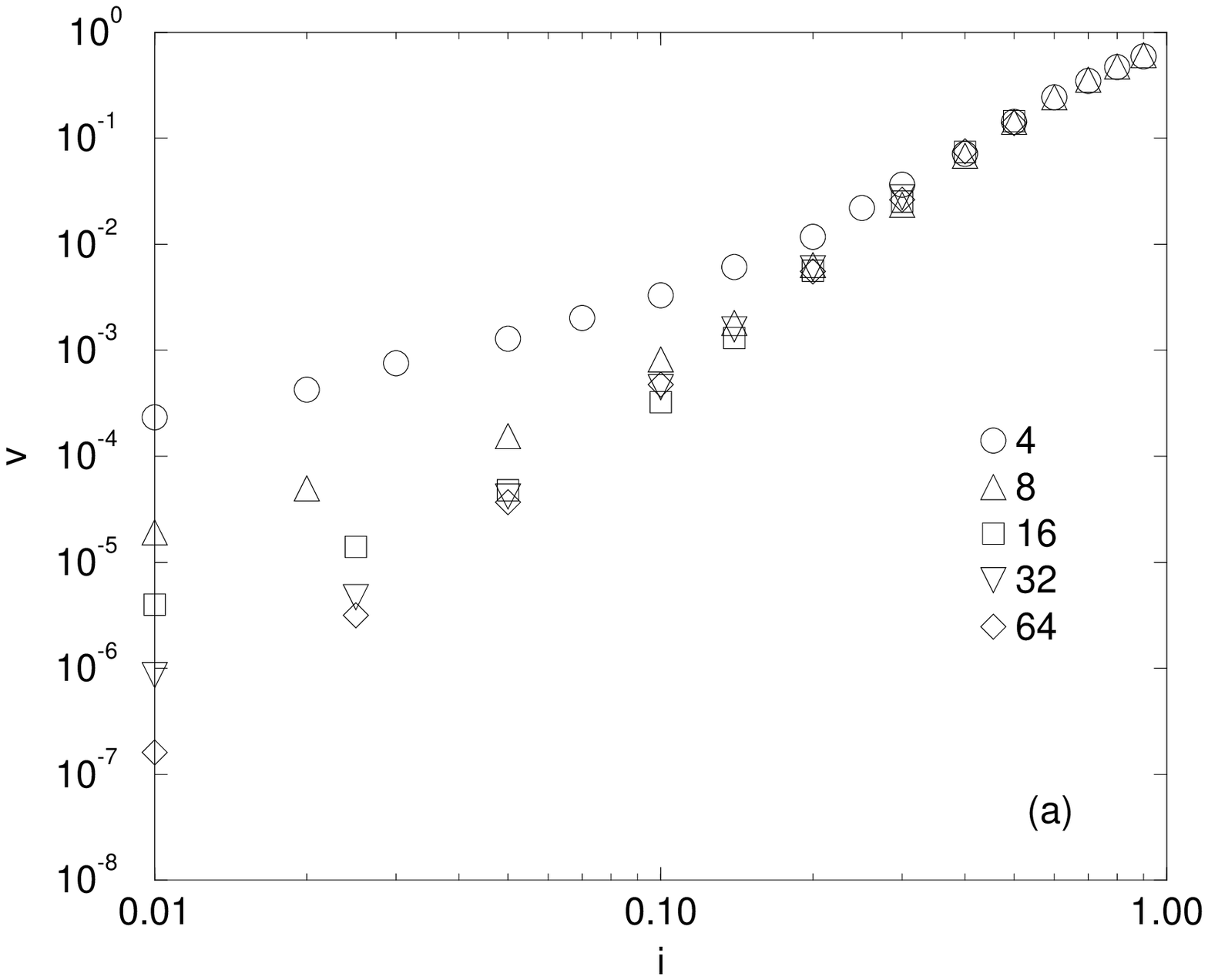}
\epsfxsize= 7 cm \epsfbox{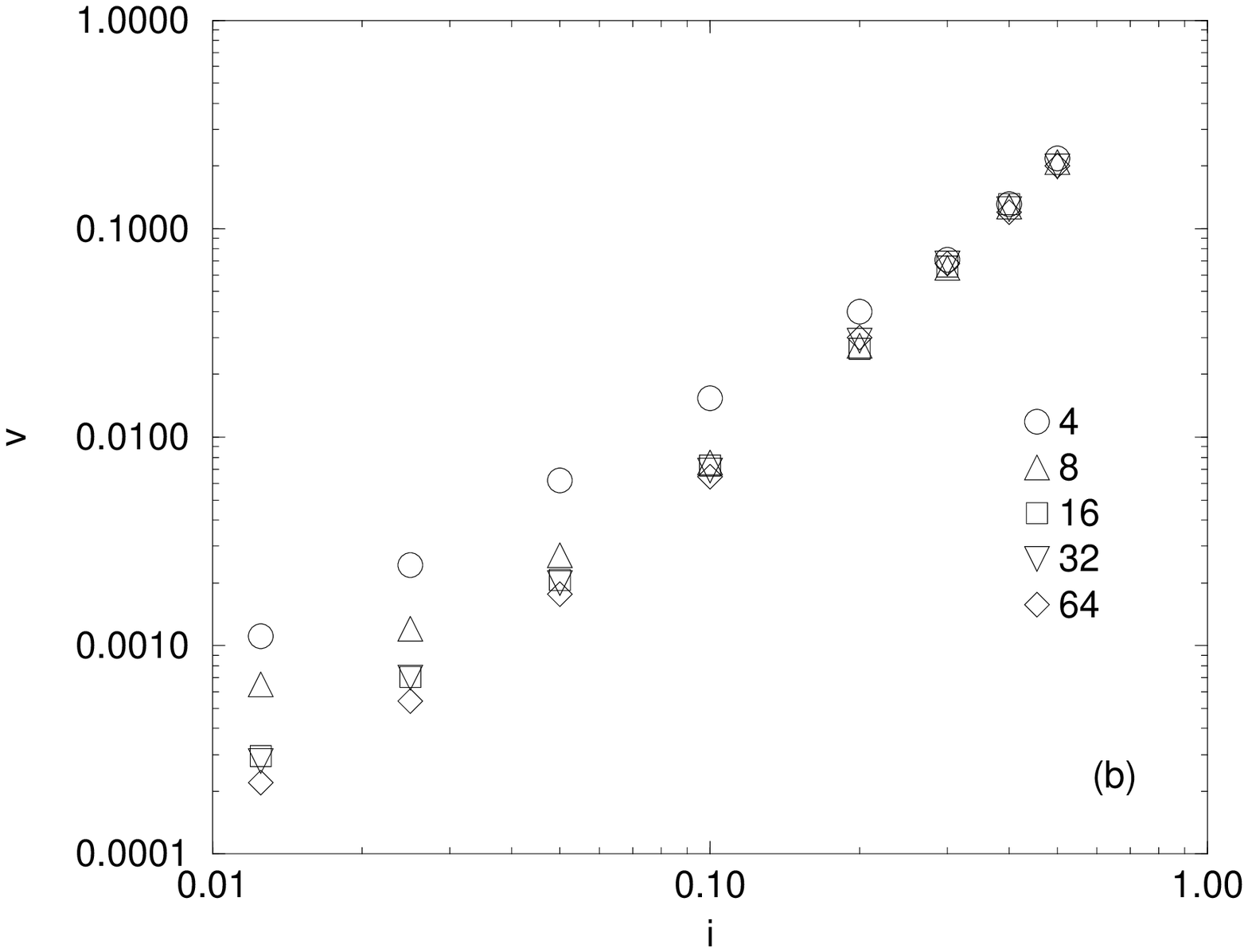}
\epsfxsize= 7 cm \epsfbox{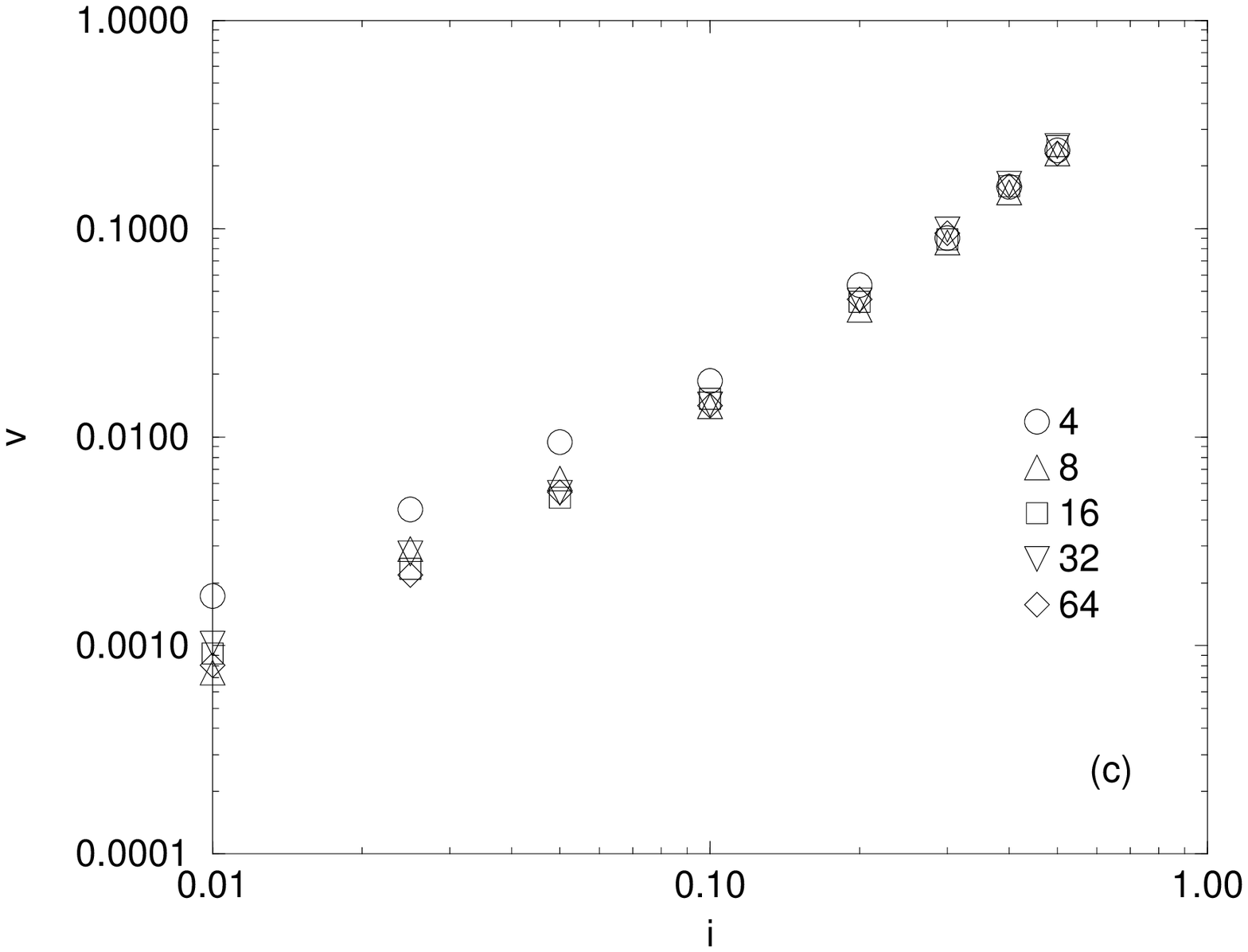}
\epsfxsize= 7 cm \epsfbox{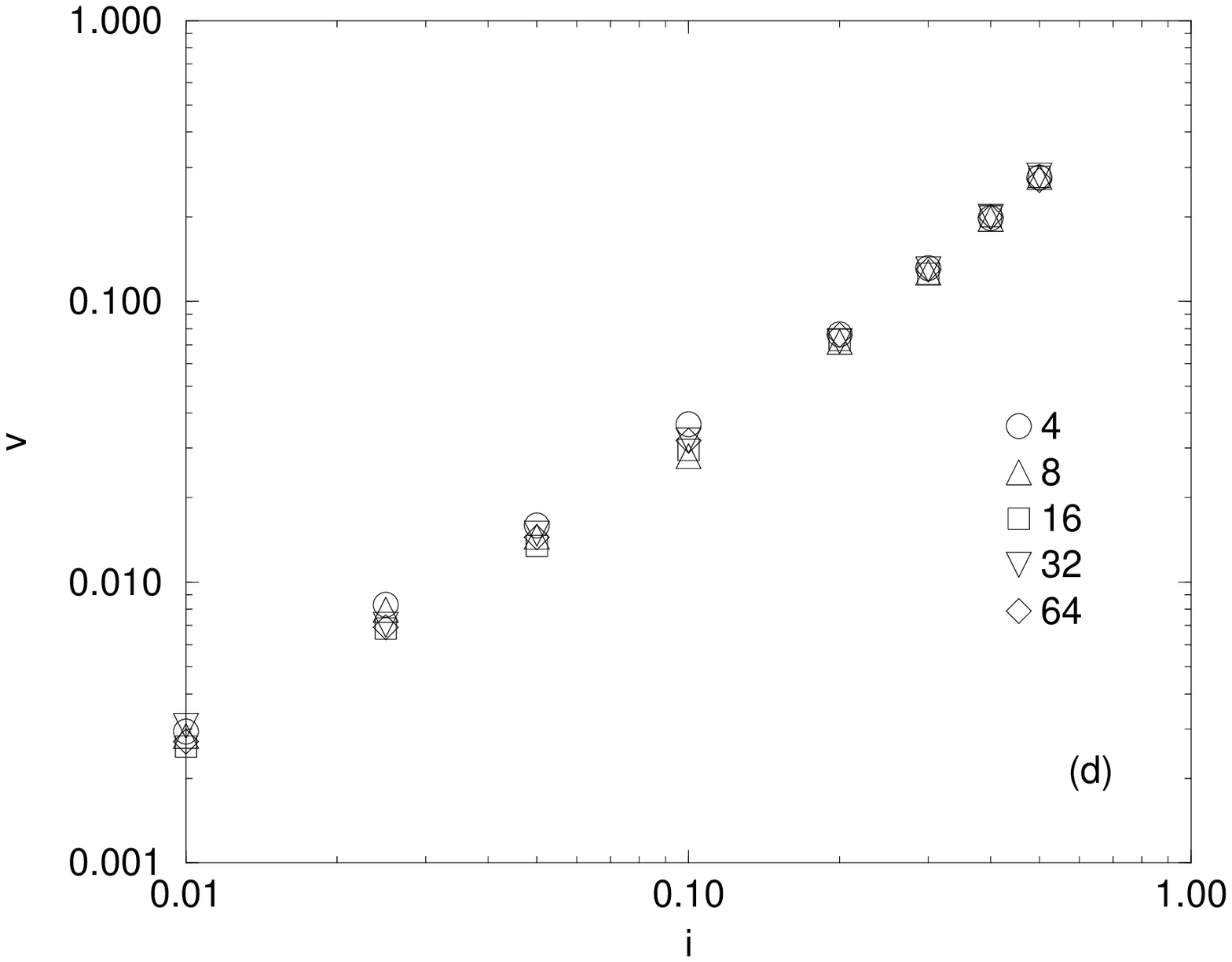}
\caption{$IV$ relation ($i=I/I_0$ versus $v=V/RI_0$) with current injection 
from busbars for arrays of
sizes $N=4,8,16,32,64$. Errors are
about the size of symbols. (a)$T=0.8$; (b)$T=1.0$; (c)$T=1.1$; (d)$T=1.3$}
\label{fig2}
\end{figure}

The results of the simulations are summarized in Figs.2(a-d) where the $IV$
relation at fixed $T$ is shown for several system sizes. In Fig.2(a) the
data
is shown for $T=0.8<T_{KT}$. At this temperature, we expect that
$\xi_{-}$ is small relative to all array sizes $N$ and to $\xi_{I}$
so that the voltage $V$ will depend mainly on the ratio
$\xi_{I}/N$ and, from our scaling hypothesis of Eq.(\ref{eq.one}), we expect
that $V\sim \xi_{I}^{-a(T)}$ when $\xi_{I}<N$ and $V\sim I$ when $\xi_{I}>N$
with a crossover between the two behaviors when $\xi_{I}/N=O(1)$.
The data of Fig.2(a) is completely consistent with
this hypothesis with the cross-over from non-linear to linear Ohmic behavior
occurs at a current $I_{x}/I_{o}=1/N$ corresponding to $\xi_{I}=N$. This
cross-over current is in good agreement with the experimental data on small
arrays: see Figs.3 and 10 of \cite{01}.
The theoretical explanation of this
cross-over to Ohmic behavior at small currents at $T<T_{KT}$ is quite
simple. The origin of the non-linear $IV$ relation in $2D$ superconductors
at $T<T_{KT}$ is that vortex pairs of separation $r>\xi_{I}=I_{o}/I$ are
unbound by the applied current $I$ at a rate $\sim(I/I_{o})^{2\pi K_{R}(T)}$
and reformed at rate $\sim n_{v}^{2}$. In the steady state, these rates are
equal so that $n_{v}\sim(I/I_{o})^{\pi K_{R}(T)}$. Since the resistivity
$V/I\propto n_{v}$, this yields the AHNS result\cite{ahns} $V\sim I^{a(T)}$
with $a(T)=\pi K_{R}(T)+1$. We note that the $IV$ relation of Fig.2a agrees
fairly well with the AHNS value for the exponent $a(T)$ at $T=0.8$ when the
theoretical estimate is $a(T)\approx 3.8$. However, our simulations are not
consistent with a recent proposal that $a(T)=2\pi K_{R}(T)-1 \approx 4.5$ at
$T=0.8$\cite{10}. If $\xi_{I}>N$, no vortex pairs are unbound by the
applied current as there are no pairs of separation $r>\xi_{I}$. so that the
only source of free vortices is by thermal excitation in the finite system
with $n_{v}\sim N^{-\pi K_{R}(T)}$ which yields a small linear resistivity
which decreases as $N$ increases.

When $T>T_{KT}$, the thermal length scale is
$\xi_{+}(\tau)\approx\xi_{-}^{2\pi}(-\tau)$\cite{ahns} where $\tau =
(T-T_{KT})/T_{KT}$
is the reduced temperature\cite{ahns}. This correlation length is much
larger than $\xi_{-}(|\tau|)$
for the same $|\tau|$ and cannot be considered
small relative to $\xi_{I}$ or to $N$ for all our arrays. At very small
applied currents $I/I_{o}<1/\xi_{+}$, the system
contains bound vortex pairs of separation $r<\xi_{+}$ and also free vortices
of areal density $n_{v}\propto\xi_{+}^{-2}$.
The dissipation will be dominated by these free
vortices leading to a linear Ohmic $IV$ relation with a resistivity $\propto
n_{v}$. However, at larger currents $\xi_{I}<\xi_{+}$ the current length
scale will control the dissipation and one expects a similar mechanism of
vortex pair unbinding and recapture as when $\tau <0$ leading to $V\sim
I^{a(T)}$ with $1<a(T)<3$. These qualitative arguments imply that the
crossover regime from the large current power law $IV$ relation to the final
linear Ohmic relation for small $I/I_{o}$ is very wide, possibly several
orders of magnitude when $T\approx T_{KT}$ and $N>\xi_{+}$, which is
consistent with experimental observation\cite{04,05}.

The simulation results for $T>T_{KT}$ are shown in Figs.2(b),(c),(d). From
these, it is clear that the cross-over from a linear to a non-linear $IV$
rlation is governed by competition between three length scales
$N,\xi_{I},\xi_{+}(T)$. For $T=1.3$ (Fig.2d) where we expect the thermal
length $\xi_{+}$ to be rather small, the $IV$ relation is independent of the
array size and the data may be interpreted by assuming $\xi_{+}<N$ for all
arrays and that the cross-over from non-linear to linear behavior is
controlled by $\xi_{+}/\xi_{I}$ with $\xi_{+}(T=1.3)\sim 4$. At the two
lower
temperatures of Figs.2(b),(c), the data may be similarly interpreted with
the added complication of additional finite size induced cross-overs. At
$T=1.0$, Fig.2(b) shows that the three largest systems with $N=16,32,64$
start to deviate from power law behavior $V\sim I^{a}$ at the same current
$I_{f}$, which we interpret as indicating that $\xi_{+}/\xi_{I}$ is the
controlling quantity. The two smaller arrays with $N=4,8$ deviate from the
others when $\xi_{I}/N>1$, roughly as they do when $T<T_{KT}$. We see both
from numerical simulations and scaling arguments that the linear $IV$
relation in superconducting arrays below $T_{KT}$ for small currents is
entirely due to finite size effects and the non-linear behavior above
$T_{KT}$ is a finite current effect. A consistent interpretation of the data
of Figs.2(b),(c),(d) for $T>T_{KT}$ may be made and we infer that the
correlation lengths $\xi_{+}(T)\sim 16,8,4$ at $T=1.0,1.1,1.3$ in fair
agreement with estimates from equilibrium simulations\cite{gb}.

\begin{figure}
\epsfxsize= 7 cm \epsfbox{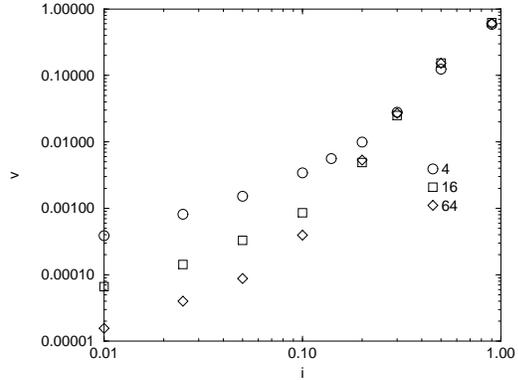}
\caption{$IV$ relation ($i=I/I_0$ versus $v=V/RI_0$) with uniform current 
injection for sizes $N=4,16,64$
at $T=0.8$.}
\label{fig3}
\end{figure}

All the simulations discussed so far were performed in the busbar geometry
and with free boundary conditions in the transverse direction to imitate the
standard experimental configuration. However, several simulations have been
done with uniform injection and extraction of the current and with periodic
boundary conditions in the transverse direction. As discussed earlier,
uniform injection is expected to emphasize dissipation at the edges of the
array. To compare with current injection from busbars, we also did
simulations at $T=0.8$ with uniform injection and transverse periodic BC for
three array sizes $N=4,16,64$ which are shown in Fig.3. Particularly for
$N=64$, the $IV$ relation is linear at much larger current than for the
busbar
geometry which implies that there is some extra dissipation. For $N=4$, the
crossover seems to occur at about the same current as in the busbar
geometry.
The explanation of this is that vortices are
readily nucleated at the edges as vortex/image pairs and, once created, are
attracted to the edge so that much of the voltage drop or dissipation occurs
not in the bulk but at the edges. This is confirmed by studying the phase
profile across the array. In Fig.4(a) is shown the phase profile for
injection from busbars. The phase is constant in thin layers adjacent to the
bars and all the phase change occurs in the bulk. Since the voltage drop
$\Delta V\propto \Delta\theta$ from Eq.(\ref{eq:v}) there is no dissipation
at the edges and it is all in the bulk. On the other hand, Fig.4(b) shows the
phase profile across the array for uniform injection. As expected, the major
part of the phase change occurs in two thin layers at the boundaries which
corresponds to motion of free vortices along the edges. This will give rise
to a linear $IV$ relation with a finite resistance which, although an edge
effect which presumably is negligible in the thermodynamic limit, will
dominate in a finite system at sufficiently low current.

\begin{figure}
\epsfxsize= 7 cm \epsfbox{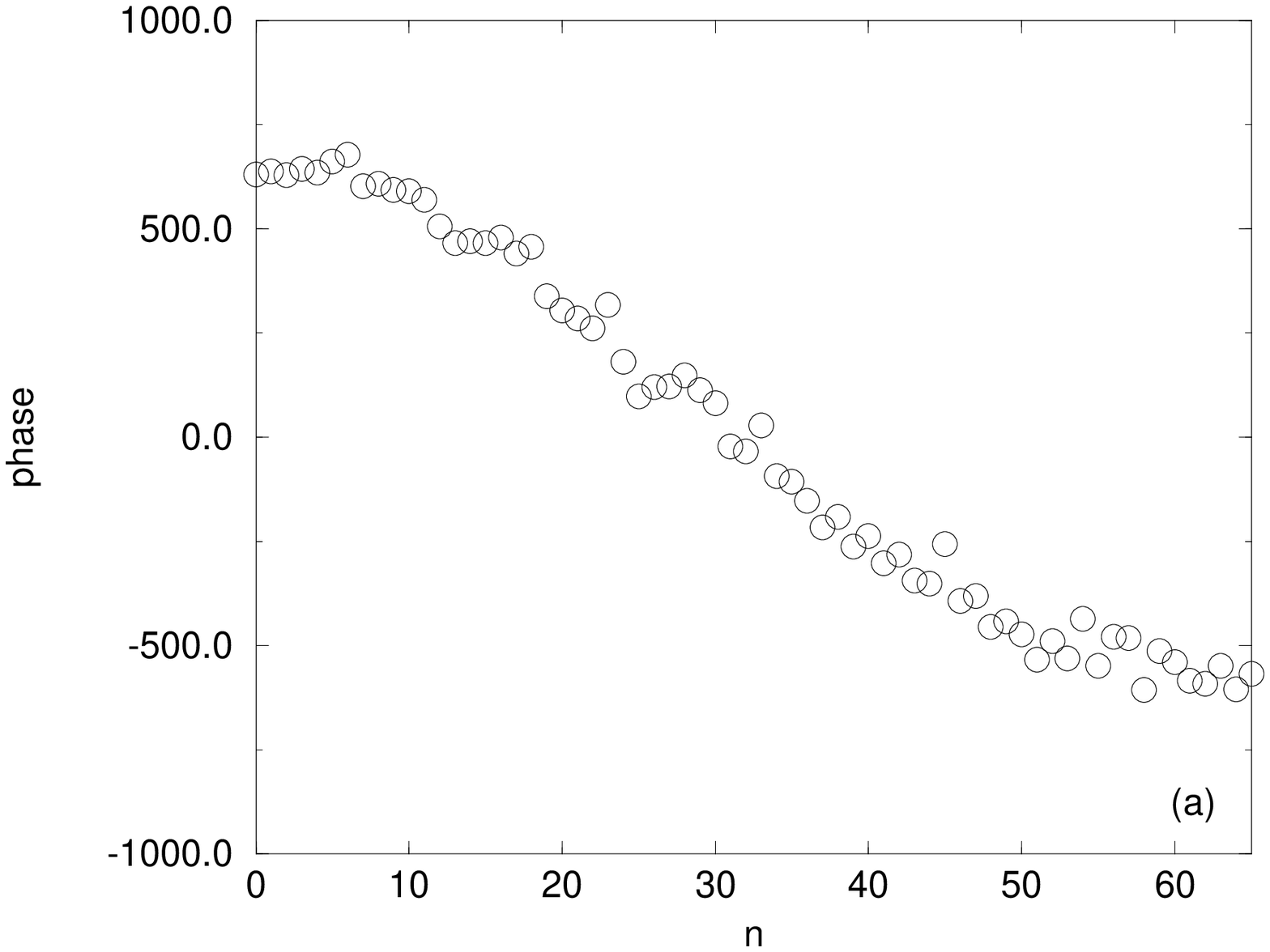}
\epsfxsize= 7 cm \epsfbox{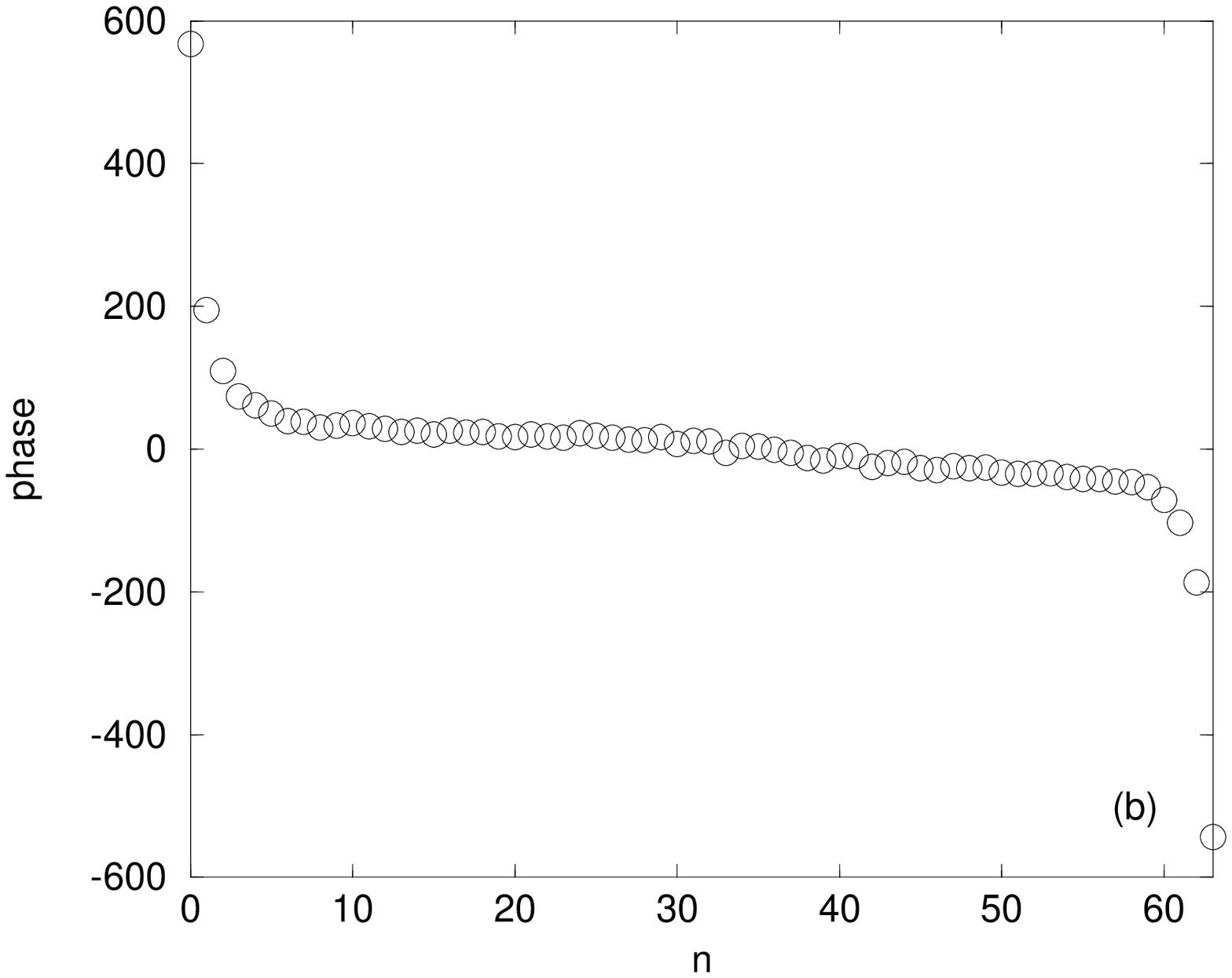}
\caption{Phase profile of array at time $t_{r}$. $N=64, I/I_{o}=0.05,
T=0.8$. The average phase of the grains in the $n^{th}$ column plotted
against column number $n$. (a)Current injection from busbars at $n=0,65$
with free BC in transverse direction; $t_{r}=5.10^{5}$. (b)Uniform
injection; $t_{r}=2.10^{5}$ and periodic BC in transverse direction.}
\label{fig4}
\end{figure}

In conclusion, this work shows that experimental and simulation studies of
the $IV$ characteristics of superconducting arrays are liable to be
dominated by finite size and edge effects unless great care is taken to
minimize them. Large array sizes seem to be essential to make any meaningful
comparison between theory and experiment or numerical simulation. When
performing a DC measurement of the $IV$ relation, dissipation at the edges
should be minimized by current injection from busbars as edge effects can
give a large contribution to the linear resistivity which may dominate the
non-linear bulk contribution. Hopefully, this paper gives some indication
how to take account of such effects and will motivate research into finding
a complete scaling form  for the voltage for finite size and finite currents
to obtain a better understanding of the complete $IV$ curves and not just
in special limits as we do at present.

Acknowledgements:
The authors are grateful to R.A. Pelcovits and J.M. Valles for
invaluable conversations. This work was supported by NSF grant DMR-9222812.
Computations were done at the Theoretical Physics Computing Facility at
Brown University.

\end{multicols}

\end{document}